\documentclass[12pt]{article}
\usepackage{amssymb,latexsym}            
\usepackage{epsf}                        

\newcommand{\sect}[1]{\setcounter{equation}{0}\section{#1}}

\def\be{\begin{equation}}
\def\ee{\end{equation}}
\def\bea{\begin{eqnarray}}
\def\eea{\end{eqnarray}}
\def\barray{\begin{array}}
\def\earray{\end{array}} 
\def\nn{\nonumber \\}
\def\vsp#1{\vspace{#1}}
\def\hsp#1{\hspace{#1}}

\def\part{\partial}

\def\tfrac#1#2{{\textstyle{#1\over #2}}}
\def\half{\tfrac{1}{2}}
\def\x{\times}

\def\Tr{\mbox{Tr}}

\def\Ovt{\ensuremath{O(4,20)\ }}
\def\Ove{\ensuremath{O(5,21)\ }}
\def\mIIA{$m$IIA}

\def\Ortin{Ort{\'\i}n}

\def\cA{\cal A}  \def\hA{\hat A}       
  \def\hB{\hat B}   
  \def\hC{\hat C}

  \def\hF{\hat F}  
    
  \def\hH{\hat H}

\def\cL{\cal L}    
\def\cM{\cal M}  \def\hM{\hat M}

  \def\hR{\hat R}

  \def\hV{\hat V}

\def\hcA{\hat{\cA}}

\def\he{\hat e}      
      
\def\hg{\hat g}

\def\hell{\hat \ell}

\def\hlambda{\hat{\lambda}}              
\def\hmu{\hat{\mu}}                  
\def\hnu{\hat{\nu}}                        
\def\hxi{\hat{\xi}}                  
                      
\def\hrho{\hat{\rho}}                
\def\hsigma{\hat{\sigma}}             
\def\htau{\hat{\tau}}                
           
\def\hphi{\hat{\phi}}                       
          
\def\hchi{\hat{\chi}}                      
                   
\def\homega{\hat{\omega}}

\def\hSigma{\hat{\Sigma}}

\def\mn{{\mu\nu}}
\def\mnr{{\mu\nu\rho}}
\def\hmn{{\hmu\hnu}}
\def\hmnr{{\hmu\hnu\hrho}}

\def\Bmn{B_{\mu\nu}}
\def\Fmn{F_{\mu\nu}}
\def\Hmnr{H_{\mu\nu\rho}}

\def\hgmn{{\hat g}_{{\hat\mu}{\hat\nu}}}
\def\hBmn{{\hat B}_{{\hat\mu}{\hat\nu}}}

\def\hHmnr{{\hat H}_{{\hat\mu}{\hat\nu}{\hat\rho}}}

\def\hVm{{\hV}_{\hmu}}

\def\epsvijf{\epsilon^{\mu\nu\rho\lambda\sigma}}
\def\epszes{\epsilon^{\hmu\hnu\hrho\hlambda\hsigma\htau}}

\def\Mab{M_{ab}^{-1}}
\def\hMab{\hM_{ab}^{-1}}

\def\hcMij{{\hat{\cM}}_{ij}^{-1}}

\def\sqrtg{\sqrt{|g|}}

\def\hsqrtg{\sqrt{|{\hat g}|}}

\def\makeatletter{\catcode`\@=11}
\makeatletter
\def\mathbox#1{\hbox{$\m@th#1$}}%
\def\math@ccstyles#1#2#3#4#5#6#7{{\leavevmode
      \setbox0\mathbox{#6#7}%
      \setbox2\mathbox{#4#5}%
      \dimen@ #3%
      \baselineskip\z@\lineskiplimit#1\lineskip\z@
      \vbox{\ialign{##\crcr
             \hfil \kern #2\box2 \hfil\crcr
             \noalign{\kern\dimen@}%
             \hfil\box0\hfil\crcr}}}}
\def\mathaccstyles{\math@ccstyles\maxdimen}
\def\maththroughstyles{\math@ccstyles{-\maxdimen}}
\def\unity%
 {\maththroughstyles{.45\ht0}\z@\displaystyle {\mathchar"006C}\displaystyle 1}
\begin{document}

\rightline{DCPT--01/43}
\rightline{hep-th/0105016}
\rightline{\today}
\vspace{2truecm}

\centerline{\Large \bf  Massive T-duality in six dimensions}
\vspace{1.3truecm}

\centerline{
    {\bf Bert Janssen}\footnote{E-mail address: 
                                  {\tt bert.janssen@durham.ac.uk}},
                                                            }

\vspace{.4cm}
\centerline{{\it Department of Mathematical Sciences}}
\centerline{{\it South Road, Durham DH1 3LE }}
\centerline{{\it United Kingdom}} 

\vspace{2truecm}

\centerline{\bf ABSTRACT}
\vspace{.5truecm}

\noindent
A massive version of T-duality in six dimensions is given, that maps 
the $K3$ compactification of Romans' theory onto the $K3$ compactification 
of Type IIB theory. This is done by performing a (standard) Kaluza-Klein 
reduction on six-dimensional massive Type IIA and a Scherk-Schwarz 
reduction on Type IIB, mapping both theories onto the same five-dimensional
theory. We also comment shortly on the difficulties arising if one 
intends to construct a massive generalisation of the six-dimensional 
string-string duality.


\newpage
\sect{Introduction}

It is generally accepted that the various string theories and their 
supergravity limits are all interrelated through T-, S-, or U-dualities, 
being different perturbative expansions in different points in the moduli 
space of the same M-theory (see for instance \cite{Witten, HT}). Many of 
these duality transformations have been explicitly constructed and most
of them, although not rigorously proven, are believed to hold.

There is, however, a class of supergravity theories that is much less 
studied than other theories, namely the massive supergravities. These are 
supergravities with one or more constant parameter with dimension of 
mass that can be interpreted as the Hodge dual of a $(D-1)$-form 
potential in $D$ dimensions \cite{PW, BRGPT}. Typically, massive gauge 
transformations give masses to some of the gauge fields by eating other 
fields.
 
Probably the best known massive supergravity theory is ten-dimensional 
massive Type IIA or Romans' theory \cite{Romans}, which is a generalisation
of standard Type IIA, adding a cosmological constant term. But there are
also other ways to obtain massive supergravities. Performing a generalised 
Scherk-Schwarz reduction \cite{SS} gives supergravity theories with mass 
parameters, provided that one gives extra dependences on the compactified 
coordinates to axionic fields. Alternatively, gauging some of the global 
symmetry transformations (i.e. giving them a linear dependence on the 
coordinates), either before or after dimensional reduction, leads to 
similar results  \cite{BRGPT, CLPST}-\cite{HLS}.      

Some work has been done on the generalisations of the duality symmetries to
massive supergravity theories. In \cite{BRGPT} it was shown that the 
T-duality between ten-dimensional Type IIA and Type IIB can be extended to 
a massive T-duality between Romans' theory and Type IIB, linking the 
Kaluza-Klein reduction of Romans' theory to the Scherk-Schwarz reduction of
Type IIB. It was also shown \cite{KKM, KM} that the Scherk-Schwarz 
reduction of the Heterotic theory on $T^d$ can be written in an 
$O(d,d+16)$ symmetric way and recently an explicitly \Ovt invariant 
formulation of the $K3$ compactification of Romans' theory was given 
\cite{HLS}. The latter results therefore indicate that these massive 
supergravity theories have the same T-duality groups as their massless 
counterparts, provided that the masses also transform under these symmetry 
groups.  

In this paper we want to make use of the results of \cite{HLS} and 
construct the massive T-duality between the $K3$ compactification of 
Romans' theory and the $K3$ compactification of Type IIB supergravity, 
using analogous techniques as in \cite{BRGPT}. We will gauge a particular 
subgroup of the \Ove symmetry group of Type IIB, that acts on 24 scalars 
and 26 vectors. Since the 24 scalars appear as axions (i.e. only via their 
derivative), Scherk-Schwarz reduction will give rise in five dimensions 
to the same number of masses and the same massive gauge transformations 
as one obtains through standard Kaluza-Klein reduction of  six-dimensional
massive Type IIA.

The paper is organised as follows: in section 2 we perform a standard 
Kaluza-Klein reduction of the massive Type IIA over $K3$, leading to an
\Ovt invariant 
massive $N=2$ supergravity in five dimensions. In section 3 we review some 
of the properties of Type IIB compactified over $K3$. The Scherk-Schwarz 
reduction of Type IIB and the obtained massive T-duality rules are given 
in sections 4 and 5 and in section 6 we comment briefly on the
impossibility of generalising the six-dimensional string-string duality
between Type IIA and Heterotic to the massive case.  In section 7 we 
summarise our conclusions.

\sect{Kaluza-Klein reduction of \mIIA/$K3$}

\noindent
The $K3$ compactification of ten-dimensional Type IIA supergravity has been
known for a long time \cite{DNP, DN}. Its bosonic field content consists 
of a six-dimensional metric $\hgmn$, a dilaton $\hphi$, a Kalb-Ramond field
$\hBmn$, 24 vectors $\hV_{\hmu}^a$ forming a vector representation of 
$O(4,20)$ and 80 scalars which parametrise an $O(4,20)/(O(4)\x O(20))$ 
coset and can be combined into a symmetric $O(4,20)$ matrix $\hMab$.
This field content arises naturally from the geometrical pro\-perties of 
the $K3$ manifold, more precisely from the 22 non-contractable two-cycles 
and the four-cycle over which the various ten-dimensional fields can 
be wrapped.
  
Recently also a $K3$ compactification of massive ten-dimensional Type IIA
supergravity has been given \cite{HLS}. Here it was 
shown that, apart from the fields mentioned above, one also generates 24 
mass parameters $m^a$, that transform in the vector representation of 
$O(4,20)$.

Twenty two of these masses come from wrapping the ten-dimensional
RR two-form field strength over the 22 two-cycles of $K3$ and one mass 
comes from wrapping the ten-dimensional RR four-form field strength over 
the four-cycle, in a kind of Scherk-Schwarz-like reduction over 
$K3$. This is possible because of the fact that the RR one- and three-form 
fields only occur via their respective field strengths (up to field 
redefinitions) \cite{LLP}. The last mass parameter is coming directly 
from the mass in ten-dimensional Romans' theory. 
It is remarkable that one has to take into account the mass coming from
Romans' theory in order to construct a full $O(4,20)$ invariant massive 
Type IIA theory in six dimensions. 
Indeed, it is claimed in \cite{HLS} that the (Scherk-Schwarz) 
compactification of Romans' 
theory over $K3$ yields a unification of various six-dimensional massive 
supergravity theories, related by $O(4,20)$ symmetry.       
  
\noindent
The bosonic part of the six-dimensional massive Type IIA action (\mIIA) is 
given by \cite{HLS}:
\bea
{\cL}_{_{\mbox{\tiny \mIIA}}} &=& \hsqrtg \Bigl \{ e^{-2\hphi} \Bigl[
            \hR\  -\  4 (\part \hphi)^2 
            \ +\ \tfrac{1}{12} \hH_{\hmnr}(\hB) \hH^{\hmnr}(\hB)
            \ +\ \tfrac{1}{8} \Tr (\part \hM^{-1} \part \hM) \Bigr]
                       \nn [.3cm]
 &&\hsp{5cm}
        -\  \tfrac{1}{4} \hF^a_{\hmn}(\hV) \hMab \hF^{b\hmn}(\hV)
        \  - \ 2m^a \hMab m^b           \Bigr\} 
\label{mIIA}  \\[.3cm]
 &+&  \tfrac{1}{16}\ \epszes \eta_{ab}
        \Bigl\{ \hF^a_\hmn(\hV) \hF^b_{\hrho\hlambda}(\hV)
                                        \hB_{\hsigma\htau}
               + 2 m^a \hF^b_\hmn(\hV)\hB_{\hrho\hlambda}\hB_{\hsigma\htau}
               + \tfrac{4}{3}
                m^a m^b\hB_\hmn\hB_{\hrho\hlambda}\hB_{\hsigma\htau}  
                         \Bigr\} \ .
\nonumber
\eea
The field strengths of the various gauge fields are defined as
\be
\hHmnr (\hB) = 3\part_{[\hmu} \hB_{\hnu\hrho]} \ , 
\hsp{2cm}
\hF^a_{\hmn}(\hV) = 2\part_{[\hmu} \hV^a_{\hnu]} + 2 m^a \hBmn \ ,
\ee 
and $\eta_{ab}$ is the invariant $O(4,20)$ metric
\be
\eta_{ab} = \left( \begin{array}{ccc}
              0 & \unity_4 & 0  \\
             \unity_4 & 0 & 0 \\
             0 & 0 & \unity_{16} 
        \end{array} \right) \ . 
\label{eta}
\ee
The gauge fields transform under the following gauge transformations
\be
\delta \hBmn = \part_{[\hmu} \hSigma_{\hnu]}\ ,
\hsp{2cm}
\delta {\hVm}^a = \part_{\hmu} \hchi^a - m^a \hSigma_{\hmu}\ .
\ee
Note that the ${\hVm}^a$ transform as a St\"uckelberg field under 
$\hSigma_{\hmu}$. Hence, by an appropriate redefinition of 
$\hBmn$, 
\be
m^a \hBmn = m^a \hB^\prime_\hmn - \part_{[\hmu} \hV^a_{\hnu]} \ , 
\ee
the Kalb-Ramond field $\hB^\prime_\hmn$ becomes massive by eating 
the vectors.

\noindent
The dimensional reduction to five dimensions is straight forward. Calling 
the direction over which we reduce $x$ and using the standard reduction 
rules\footnote{Here we denote six-dimensional fields as hatted and 
five-dimensional fields as unhatted.}  
\be
\begin{array}{ll}
\hg_{xx} = -k^{2} \ ,                  
       & \hB_{\mu x} = B_\mu \ , \\[.3cm]              
\hg_{x\mu} = -k^{2} A_\mu \ ,          
       & \hB_{\mn} = \Bmn - A_{[\mu} B_{\nu]}\ ,  \\[.3cm]
\hg_{\mn} = g_{\mn} -k^{2}A_\mu A_\nu \ ,  \hsp{2cm}
       &  \hV^a_x = \ell^a \ ,\\[.3cm]
\hphi = \phi +\half \log k \ ,
       & \hV^a_\mu =  V^a_\mu + \ell^a A_\mu \ , \\ [.3cm]
\hMab = \Mab \nonumber\ ,
       & 
\end{array}
\label{mIIAred}
\ee
we find for the five-dimensional action
\bea
{\cL}_{\mbox{\tiny $m$II}} &=& \sqrtg \Bigl\{ e^{-2\phi} \Bigl[
            R\ -\ 4 (\part \phi)^2\ +\ \tfrac{1}{12} H_\mnr(B)H^\mnr(B) 
         \ +\ \tfrac{1}{8} \Tr (\part M^{-1} \part M) \nn  [.3cm]
&& \hsp{3cm}
          + \tfrac{1}{k^2}(\part k)^2 
         \ -\ \tfrac{1}{4}k^2 F_{\mn}(A)F^{\mn}(A)
         \ -\ \tfrac{1}{4k^2} F_{\mn}(B) F^{\mn}(B) \Bigr]    \nn[.3cm]
&&  \hsp{1cm}
         - \ \tfrac{1}{4} k F_{\mn}^a(V)\Mab  F^{b\mn}(V)
        \ + \ \tfrac{1}{2k} D_\mu\ell^a \Mab D^\mu\ell^b 
         \ - \ 2 k\ m^a \Mab m^b  \Bigl\} \label{mII}\\  [.3cm]      
&+& \tfrac{1}{8}\ \epsvijf \eta_{ab} 
     \Bigl\{ \Fmn^a(V) F_{\rho\lambda}^b(V) B_\sigma
          \ + \ 2 D_\mu \ell^a F_{\nu\rho}^b(V) B_{\lambda\sigma}
          \ - \ 2  D_\mu \ell^a F_{\nu\rho}^b(V) A_\lambda B_\sigma
\nn[.3cm]
&&   \hsp{5cm}
         \ + \  4 m^a \Fmn^b(V) B_{\rho \lambda} B_\sigma 
         \  +\ 2 m^a D_\mu \ell^b B_{\nu\rho} B_{\lambda\sigma}
\nn[.3cm]
&& \hsp{5cm}
          \ -\ 4 m^a D_\mu \ell^b B_{\nu\rho}A_\lambda B_\sigma
          \ +\ 4 m^a m^b\Bmn B_{\rho \lambda} B_\sigma
\Bigr\} \ . \nonumber
\eea
Now the field strengths are given by
\bea
&&F_\mn(A) = 2\part_{[\mu} A_{\nu]} \ , \hsp{1.5cm} 
F_\mn(B) = 2\part_{[\mu} B_{\nu]}  \ , \nn[.3cm]
&&\Hmnr(B) = 3\part_{[\mu} B_{\nu\rho]} 
                 - \tfrac{3}{2}A_{[\mu}F_{\nu\rho]}(B)
                 - \tfrac{3}{2}B_{[\mu}F_{\nu\rho]}(A) 
\label{5dfstr}\\[.3cm]
&&F_\mn^a(V) = 2\part_{[\mu} V^a_{\nu]} + 2m^a \Bmn 
                 + 2 m^a A_{[\mu} B_{\nu]}
                 + 2 \ell^a \part_{[\mu} A_{\nu]} \ , \nn[.3cm]
&& D_\mu \ell^a = \part_\mu \ell^a + 2 m^a B_\mu \ , \nonumber
\eea
and can easily be checked to be invariant under the five-dimensional 
gauge transformations
\bea
&&\delta A_\mu = \part_\mu \xi  \ , \hsp{1.5cm} 
\delta B_\mu = \part_\mu \Sigma  \ ,\nn[.3cm]
&&\delta \Bmn = \part_{[\mu} \Sigma_{\nu]} 
                 + A_{[\mu}\part_{\nu]} \Sigma 
                 + B_{[\mu}\part_{\nu]} \xi\ , \nn[.3cm]
&& \delta V_\mu^a = \part_{\mu} \chi^a - m^a \Sigma_{\mu} 
                        + 2m^a\Sigma A_\mu 
\label{5dgauge} \\[.3cm]
&& \delta \ell^a = -2 m^a \Sigma\ . \nonumber
\eea
Note that the symmetry group of the reduced action is $O(4,20)$, 
instead of an $O(5,21)$ symmetry expected from the massless case 
\cite{MS}. This is of course due to the massive gauge transformations 
(\ref{5dgauge}), which break the $O(5,21)$ into $O(4,20)$, by making  
$V_\mu^a$ transform as St\"uckelberg fields on one hand, 
while on the other hand the winding vector $B_\mu$ becomes massive by 
eating the $\ell^a$'s and the Kaluza-Klein vector $A_\mu$ is inert under 
the St\"uckelberg transformations. 

In the next sections we will perform a Scherk-Schwarz reduction on 
Type IIB theory compactified over $K3$ and show that it leads to the 
same action as the one we derived above.

\sect{Type IIB on $K3$: a review} 
\noindent
The six-dimensional Type IIB supergravity is obtained via a 
compactification of the ten-dimensional Type IIB over $K3$ \cite{Townsend, 
Romans2}. Its field content consists of a metric $\hgmn$, 5 self-dual and 
21 anti-self-dual two-forms ${\hcA}^i_{\hmn}$ transforming under the 
vector representation 
of $O(5,21)$ and 105 scalars, parametrising an $O(5,21)/(O(5)\x O(21))$ 
coset. The scalars can be combined into a symmetric $O(5,21)$ matrix 
$\hcMij$. 

Due to the (anti-)self-duality conditions, it is impossible to write down 
a covariant action \cite{MaS}. However it is consistent to write down a 
non-self-dual action, dropping the self-duality condition. This 
non-self-dual action will give the same equations of motion, provided that 
one inserts the duality condition by hand as an extra equation of motion
\cite{BBO}.

\noindent
The bosonic non-self-dual action is then given by
\bea
{\cL}_{\mbox{\tiny IIB}} = \hsqrtg \Bigl\{
        \hR + \tfrac{1}{8} \Tr (\part {\hat{\cM}}^{-1} \part {\hat{\cM}}) 
            + \tfrac{1}{24} \hH^i_{\hmu\hnu\hrho} (\hcA)\hcMij
                             \hH^{j\hmu\hnu\hrho} (\hcA)\Bigr\} \ ,
\label{IIBE}
\eea 
where the three-form field strength
\be
\hH^i_{\hmu\hnu\hrho} (\hcA) = 3 \part_{[\hmu} \hcA^i_{\hnu\hrho]} 
\ee    
satisfies the (anti-)self-duality condition
\be
\hH^i_{\hmu\hnu\hrho} (\hcA) = \eta^{ij}
          \hat{\cM}^{-1}_{jk}\  {}^*\hH^k_{\hmu\hnu\hrho} (\hcA) \ .
\label{self-dual6}
\ee
Here $\eta^{ij}$ is the invariant $O(5,21)$ metric
\be
\eta_{ij} = \left( \begin{array}{ccc}
              0 & 1 & 0  \\
              1 & 0 & 0 \\
              0 & 0 & \eta_{ab} 
        \end{array} \right) \ , 
\ee
with $\eta_{ab}$ as defined in (\ref{eta}).

\noindent
In order to write the action (\ref{IIBE}) in the string frame, one has to 
identify which of the 105 scalars corresponds to the dilaton. This can be
done by choosing a particular parametrisation of $\hcMij$, writing the 105 
$O(5,21)$ scalars as 80 scalars contained in an $O(4,20)$ matrix $\hMab$, 
24 scalars $\hell^a$ transforming as a vector under $O(4,20)$ and the 
dilaton $\hphi$:
\begin{equation}
{\hat {\cal M}}^{-1}_{ij} =
\left(\begin{array}{ccc}
e^{-2{\hat \phi}} + {\hat \ell}^a  {\hat { M}}^{-1}_{ab}{\hat \ell}^b
+{1\over 4}e^{2{\hat \phi}}{\hat\ell}^4&
-{1\over 2}e^{2{\hat\phi}}{\hat\ell}^2&
{\hat\ell}^a{\hat M}^{-1}_{ab} + {1\over 2}e^{2{\hat\phi}}{\hat\ell}^2
{\hat\ell}^a \eta_{ab}\\ [.3cm]
-{1\over 2}e^{2{\hat\phi}}{\hat\ell}^2&
e^{2\hat\phi}&
-e^{2{\hat\phi}}{\hat\ell}^a \eta_{ab} \\ [.3cm]
{\hat M}^{-1}_{ab}{\hat\ell}^b + {1\over 2}e^{2{\hat\phi}}{\hat\ell}^2
 \eta_{ab}{\hat\ell}^b &
-e^{2{\hat\phi}} \eta_{ab}{\hat\ell}^b &
{\hat M}_{ab}^{-1} + e^{2\hat\phi}{\hat\ell}^c{\hat\ell}^d 
\eta_{ac}\eta_{bd}
\end{array}\right) \ ,
\label{matrix}
\end{equation}
where with ${\hell}^2$ we mean $\hell^2 = \hell^a \hell^b \eta_{ab}$. 

\noindent

Hence the action (\ref{IIBE}) in the string frame is give by \cite{BBJ}
\bea
{\cL}_{\mbox{\tiny IIB}} = \hsqrtg \Bigl\{ e^{-2\hphi} \Bigl[
                 \hR - 4 (\part \phi)^2 
                 + \tfrac{1}{8} \Tr (\part\hM^{-1} \part \hM) \Bigr] 
                 +\half \part\hell^a \hMab\part\hell^b
            + \tfrac{1}{24} \hH^i \hcMij \hH^{j}\Bigr\}. &&
\label{IIBs}
\eea  
Note that by choosing this particular parametrisation of $\hcMij$, we are 
forced to write the action (\ref{IIBs}) in a manifest \Ovt invariant way, 
though of course it is still implicit \Ove symmetric. Let us for later 
convenience also denote the \Ove vector $\hcA^i_{\hmu\hnu}$ and its gauge 
transformations in an explicit \Ovt symmetric way:
\be
\hcA^i_{\hmu\hnu}  = 
     \left( \begin{array}{c}
              \hA_{\hmu\hnu}  \\[.3cm]
              \hC_{\hmu\hnu}   \\[.3cm]
              \hV^a_{\hmu\hnu}  
           \end{array} \right) \ ,
\hsp{1.5cm}
\delta \hcA^i_{\hmu\hnu}  = 
     \left( \begin{array}{c}
              \part_{[\hmu}\hxi_{\hnu]}  \\[.3cm]
              \part_{[\hmu}\homega_{\hnu]}   \\[.3cm]
              \part_{[\hmu}\hchi^a_{\hnu]}  
           \end{array} \right) \ .
\label{2vorm}
\ee  

\sect{Scherk-Schwarz reduction of Type IIB}
There does not exist a  massive version of ten-dimensional Type IIB 
supergravity, nor is it possible to make a Scherk-Schwarz-like reduction 
over $K3$. Yet, it is still possible to relate this theory to \mIIA\ 
reduced over $K3$, generalising the well known T-duality between massless 
Type IIA and Type IIB in six dimensions \cite{Witten, DLR}. This can be 
done by performing a 
Scherk-Schwarz reduction of six-dimensional Type IIB and mapping it onto 
the five-dimensional action (\ref{mII}) of \mIIA. In this way we can 
construct the massive T-duality rules, completely analogous to the massive 
T-duality rules between Romans' theory and Type IIB supergravity in ten 
dimensions \cite{BRGPT}.

In practice, we will use the techniques explained in \cite{BRGPT, CLPST,
LLP, BRE}. The idea consists in interpreting the five-dimensional 
St\"uckelberg transformations (\ref{5dgauge}) on the Type IIB side as 
general coordinate transformations in the compactified $x$-direction 
\cite{BRGPT}. Giving an extra $x$-dependence to the fields will generate 
mass terms and St\"uckelberg transformations after reduction.

Alternatively, one can also take an \Ove symmetry that shifts 
the 24 scalars $\hell^a \rightarrow \hell^a + \Lambda^a$ and gauge this 
transformation,  giving $\Lambda^a$ a linear $x$-dependence \cite{BRE}. The
slope parameters in this linear dependence appear as 24 masses after 
reduction. This particular \Ove transformation that generates the shifts, 
is given by \cite{BJO, BRE}:
\be
\homega(\Lambda) = 
 \left( \begin{array}{ccc}
             1& 0 & 0   \\
             -\half \Lambda^2 & 1 & \Lambda^a\eta_{ab}  \\
             - \Lambda^a & 0 & \delta^a{}_b 
           \end{array} \right) \ ,
\label{gauge}
\ee
and acts on the scalars $\hell^a$ and the two-forms $\hcA_\hmn$ as follows:
\bea
&& \hell^a \rightarrow \hell^a + \Lambda^a \nn
&& \hA_{\hmu\hnu}  \rightarrow \hA_{\hmu\hnu}  \nn
&& \hC_{\hmu\hnu}  \rightarrow \hC_{\hmu\hnu} 
                   + \Lambda^a \hV^b_{\hmu\hnu} \eta_{ab}    
                   -\half \Lambda^2 \hA_{\hmu\hnu}\\
&& \hV^a_{\hmu\hnu} \rightarrow \hV^a_{\hmu\hnu} - \Lambda^a\hA_{\hmu\hnu}
\nonumber
\eea
In order to match the action (\ref{IIBs}) onto (\ref{mII}), we have to 
chose $\Lambda^a(x) = -2m^a x$ and use the following reduction rules:
\be
\begin{array}{ll}
\hg_{xx} = -k^{-2} \ ,                  
       & \hA_{\mu x} = A_\mu \ , \\ [.3cm]             
\hg_{x\mu} = -k^{-2} B_\mu \ ,          
       & \hA_{\mn} = \Bmn - B_{[\mu} A_{\nu]}\ ,  \\[.3cm]
\hg_{\mn} = g_{\mn} -k^{-2}B_\mu B_\nu \ ,  \hsp{1.5cm}
       &  \hV^a_{\mu x} = V^a_\mu + 2m^ax A_\mu  \ ,\\[.3cm]
\hphi = \phi -\half \log k \ ,
       & \hV^a_\mn =  V^a_\mn - B_{[\mu} V^a_{\nu]} 
                              +  2m^a x [\Bmn - B_{[\mu} A_{\nu]} ] 
                                                 \\ [.3cm]
\hMab = \Mab \ ,
       & \hC_{\mu x}= C_\mu - 2m^a x V^a_\mu \eta_{ab} 
                             - 2 m^2 x^2 A_\mu\\ [.3cm]
\hell^a = \ell^a -2 m^a x 
       & \hC_{\mn}= C_\mn - B_{[\mu} C_{\nu]} 
                      - 2m^a x [V^b_\mn - B_{[\mu} V^b_{\nu]}]\eta_{ab} 
             \\[.3cm]
       & \hsp{2cm} 
         - 2 m^2 x^2 [ \Bmn - B_{[\mu} A_{\nu]}] 
\end{array}
\label{IIBred}
\ee
Note that, apart from the explicit $x$-dependence, the reduction rules 
given above vary from the ones in (\ref{mIIAred}). In fact we are using 
the so-called T-dual version of (\ref{mIIAred}), in order to map the two
actions directly onto the same action in five-dimensions. This technique 
was used for example in \cite{BHO}. 

It is not difficult to see that with this reduction Ansatz the kinetic 
term for the scalars $\hell^a$ reduces in the right way and gives in five 
dimensions an $x$-independent result: \footnote{To work out
the reduction of the field strengths it is convenient to go to Vielbein 
notation $\part_{\hA} \hell^a= \he^{\hmu}_{\hA}\ \part_{\hmu} \hell^a $ 
and use the following reduction Ansatz: 
$\he^{\hmu}_{\hA}= 
\left( \begin{array}{cc}
             e^\mu_A& -B_A    \\
             0 & k 
           \end{array} \right)
$.} 
\be
\half \part_{\hA} \hell^a \hMab \part^{\hA} \hell^b =
\half D_{A} \ell^a \Mab D^{A} \ell^b - 2 k^2 m^a \Mab m^b \ .
\ee 
Note that we generated naturally a covariant derivative for $\ell^a$, 
invariant under the St\"u\-ckel\-berg transformations (\ref{5dgauge}):
\be
D_{\mu} \ell^a = \part_\mu \ell^a + 2m^a B_\mu \ .
\ee 
The naive reduction of the field strengths $\hH^i_{\hmu\hnu\hrho} (\hcA)$,
however, does give $x$-dependent results. For example,
\be
\hH^a_{ABX} (\hV) = \he^\mu_A\he^\nu_B\he^x_X \Bigl[
                    2 \part_{[\mu} V^a_{\nu]} 
                    + 4 m^a x \part_{[\mu} A_{\nu]}
                    + 2 m^a \Bmn - 2m^a B_{[\mu} A_{\nu]} \Bigr]\ .
\ee  
Hence, instead of identifying $\hH^a_{ABX} (\hV)$ with the 
five-dimensional field strength  $F_{AB}^a(V)$, as one does in the case 
of standard Kaluza-Klein 
reduction, we define the five-dimensional field strength as 
\bea
k F^a_{AB}(V) &\equiv& \hH^a_{ABX} (\hV) + \hell^a \hH_{ABX} (\hA) 
                                                \nn [.3cm]
              &=& e^\mu_A e^\nu_B \Bigl[
                   2 \part_{[\mu} V^a_{\nu]} 
                   + 2 m^a \Bmn - 2m^a B_{[\mu} A_{\nu]} 
                   + 2 \ell^a \part_{[\mu} A_{\nu]}\Bigl] \ .
\label{fstrred}
\eea
This definition of $F^a_\mn(V)$ is not only $x$-independent, but also 
coincides with the definition 
(\ref{5dfstr}) and is therefore invariant under the gauge transformations
(\ref{5dgauge}). Similarly for the other field strengths:
\bea
H_{ABC}^a (V) &\equiv& \hH^a_{ABC} (\hV) + \hell^a \hH_{ABC} (\hA)\ , 
\nn[.3cm]
k F_{AB} (C) &\equiv& \hH^a_{ABX} (\hC) 
                - \hell^a \hH_{ABX}^b (\hV)\eta_{ab}
                  + \half \hell^a  \hH_{ABX} (\hA) \ , 
\label{fstrred2}\\[.3cm]
H_{ABC} (C) &\equiv& \hH^a_{ABC} (\hC) - \hell^a \hH_{ABC}^b (\hV)\eta_{ab}
                  + \half \hell^2  \hH_{ABC} (\hA)\ .\nonumber
\eea
Hence, we find for the five-dimensional field strengths of the reduced 
Type IIB theory:  
\bea
&&F_\mn(A) = 2\part_{[\mu} A_{\nu]} \ , \nn [.3cm]
&&F_\mn(B) = 2\part_{[\mu} B_{\nu]}  \ , \nn [.3cm]
&&\Hmnr(B) = 3\part_{[\mu} B_{\nu\rho]} 
                 -  3 A_{[\mu}\part_\nu B_{\rho]}
                 -   3 B_{[\mu}\part_\nu A_{\rho]}  \ ,   \nn[.3cm]
&&F_\mn^a(V) = 2\part_{[\mu} V^a_{\nu]} + 2m^a \Bmn 
                 + 2 m^a A_{[\mu} B_{\nu]}
                 + \ell^a \Fmn(A) \ , \nn[.3cm]
&&\Hmnr^a (V)=  3 \part_{[\mu} V^a_{\nu\rho]} 
               - 3 B_{[\mu}\part_\nu V_{\rho]}^a 
               - 3 V^a_{[\mu}\part_\nu B_{\rho]}  
               - 6m^a B_{[\mu}B_{\nu\rho]} 
               + \ell^a \Hmnr (B)   \ ,    
\label{5dfstrB}             
                                         \\ [.3cm]
&&F_\mn(C) = 2\part_{[\mu} C_{\nu]} - 2m^a V^b_\mn \eta_{ab} 
              + 2 m^a B_{[\mu} V_{\nu]}^b \eta_{ab}       
              -\ell^a \Fmn^b(V)\eta_{ab} + \half \ell^2 \Fmn(A)\ ,      
\nn[.3cm]
&&\Hmnr(C) = 3\part_{[\mu} C_{\nu\rho]} 
              - 3 C_{[\mu}\part_\nu B_{\rho]}
              - 3 B_{[\mu}\part_\nu C_{\rho]} 
              + 6m^a B_{[\mu}V^b_{\nu\rho]} \eta_{ab}  \nn 
     && \hsp{4cm} 
              -\ell^a \Hmnr^b (V)\eta_{ab}
              + \half \ell^2 \Hmnr (B) \nonumber \ . \nonumber
\eea
It can be shown that these field strengths are invariant under the 
five-dimensional gauge transformations: 
\vsp{.3cm}
\bea
&&\delta A_\mu = \part_\mu \xi  \ , \hsp{1cm} 
\delta B_\mu = \part_\mu \Sigma \ , \hsp{1cm} 
\delta \ell^a = -2 m^a \Sigma\ , \nn[.3cm]
&&\delta \Bmn = \part_{[\mu} \Sigma_{\nu]} 
                 + A_{[\mu}\part_{\nu]} \Sigma 
                 + B_{[\mu}\part_{\nu]} \xi\ , \nn [.3cm]
&& \delta V_\mu^a = \part_{\mu} \chi^a - m^a \Sigma_{\mu} 
                        + 2m^a\Sigma A_\mu\ ,  \nn [.3cm]
&& \delta C_\mu = \part_{\mu} \omega + m^a \chi^b_{\mu} \eta_{ab} 
                        - 2m^a\Sigma V^b_\mu \eta_{ab} \ ,  
\label{5dgaugeB} \\ [.3cm]
&& \delta V_\mn^a = \part_{[\mu} \chi^a_{\nu]} 
                 + V^a_{[\mu} \part_{\nu]} \Sigma 
                 + B_{[\mu}\part_{\nu]} \chi^a
                 + 2m^a\Sigma \Bmn
                 - m^a B_{[\mu} \Sigma_{\nu]}\ , \nn [.3cm]
&& \delta C_\mn = \part_{[\mu} \omega_{\nu]} 
                 + C_{[\mu} \part_{\nu]} \Sigma 
                 + B_{[\mu}\part_{\nu]} \omega
                 - 2m^a\Sigma V_\mn^b \eta_{ab}
                 + m^a B_{[\mu} \chi^b_{\nu]} \eta_{ab} \ ,
\nonumber
\eea
where we have used the following reduction rules for 
parameters of the the gauge transformations (\ref{2vorm}):
\be
\vsp{.3cm}
\begin{array}{lll}
\hxi_x = 2\xi \ , \hsp{.9cm}&
\hchi^a_x = \chi^a + 4m^a x \xi\ , \hsp{.9cm} &
\homega_x = 2 \omega -4m^a x \chi^b \eta_{ab} - 4 m^2 x^2\xi\ ,
\\[.3cm]
\hxi_\mu = \Sigma_\mu \ , &
\hchi^a_\mu=2 \chi^a_\mu + 2m^a x \Sigma_\mu\ ,  &
\homega_\mu = \omega_\mu -2m^a x \chi^b_\mu \eta_{ab} 
              - 2 m^2 x^2\Sigma_\mu \ .
\end{array}
\ee
With the definitions (\ref{fstrred})-(\ref{5dfstrB}) of the 
five-dimensional field strengths and the parametrisation (\ref{matrix}) of 
$\hcMij$, we see that reduction of the kinetic term of the two-forms 
yields:
\bea
&&\hH^i (\hcA)\ \hcMij\  \hH^{j} (\hcA) = \nn
&&\hsp{1cm} 
  = e^{-2\phi}\hH^2(\hA)  
  \ +\ \hMab \Bigl[\hH^a(\hV) + \hell^a \hH(\hA)\Bigr]
           \Bigl[\hH^b(\hV) + \hell^b \hH(\hA)\Bigr] \nn 
&& \hsp{5cm} 
    + e^{2\phi} \Bigr[ \hH(\hC) - \hell^a \hH^b(\hV) \eta_{ab} 
                              + \half \hell^2 \hH(\hA) \Bigr]^2
\\
&& \hsp{1cm} 
  =  e^{-2\phi} k \Bigl[ H^2(B) -3k^2 F^2(A) \Bigr]
    + \Mab \Bigl[ H^a(V) H^b(V) -3k^2 F^a(V) F^b(V) \Bigr] \nn
&&\hsp{5cm}
+e^{2\phi} k^{-1}  \Bigl[ H^2(C) -3k^2 F^2(C) \Bigr]\ , \nonumber
\eea
such that the full five-dimensional action (\ref{IIBs}) yields
\bea
{\cL}^\prime_{\mbox{\tiny $m$II}} &=& \sqrtg \Bigl\{ e^{-2\phi} \Bigl[
          R - 4 (\part \phi)^2 +\tfrac{1}{24} H_{\mnr}(B)H^{\mnr}(B) 
          + \tfrac{1}{8} \Tr (\part M^{-1} \part M) \nn [.3cm] 
&& \hsp{3cm}
          + \tfrac{1}{k^2}(\part k)^2 
          - \tfrac{1}{8}k^2 F_{\mn}(A) F^{\mn}(A)
          - \tfrac{1}{4k^2} F_{\mn}(B)F^{\mn}(B) \Bigr]   \nn[.3cm]
&& \hsp{1.3cm}   
          +\tfrac{1}{24k}  H^a_{\mnr}(V)\Mab  H^{b\mnr}(V)
          -\tfrac{1}{8} k F^a_{\mn}(V)\Mab  F^{b\mn}(V)   
\label{mII2}\\[.3cm]
&& \hsp{1.3cm}   
          +\tfrac{1}{24 k^2}  e^{2\phi} H_{\mnr}(C)H^{\mnr}(C)
           -\tfrac{1}{8}  e^{2\phi} F_{\mn}(C) F^{\mn}(C)  \nn[.3cm]
&& \hsp{1.3cm}   
          + \tfrac{1}{2k} D_\mu\ell^a \Mab D^\mu\ell^b
          -2 k \ m^a \Mab m^b  \Bigl\} \ .\nonumber
\eea
Note that the Scherk-Schwarz reduction has broken the \Ove symmetry into 
an $O(4,20)$, which is of course the group left invariant under the 
gauging (\ref{gauge}). Although the group structure of action (\ref{mII2})
coincides with the one from action (\ref{mII}), the actions themselves are 
clearly different. Yet it turns out that they are equivalent and will 
become identical upon dualising some of the gauge fields.

\noindent
As mentioned above, the six-dimensional gauge fields obey the 
(anti-)self-duality relation (\ref{self-dual6}), which forms an essential 
part of the theory, though not of the action. These self-duality relations 
have to be taken into account in the reduction to five dimensions, where 
they give a Hodge duality relation between the two-forms and three-forms.  
Using the reduction Ansatz (\ref{fstrred})-(\ref{5dfstrB}) and the 
parametrisation
(\ref{matrix}), we find the following five-dimensional duality relations:
\bea
&& H^\mnr (B) = \frac{1}{2\sqrtg} \ e^{2\phi} \
                   \epsvijf F_{\lambda \sigma}(C) \ , \nn[.3cm]
&& H^\mnr (C) =-\frac{1}{2\sqrtg} \ e^{-2\phi} \ k^2\ 
                   \epsvijf F_{\lambda \sigma}(A) \ , 
\label{self-dual5} \\ [.3cm]
&& H^{\mnr a} (V) =-\frac{1}{2\sqrtg} \  k \ \eta^{ab} M_{bc}^{-1}
                   \epsvijf F^c_{\lambda \sigma}(V) \ , \nonumber
\eea 
Again we see that the \Ove symmetry of (\ref{self-dual6}) gets broken 
into an \Ovt \ symmetry. Using these duality relations, we can substitute
$H^a_\mnr(V)$, $F_\mn(C)$ and $H_\mnr(C)$ in the action (\ref{mII2}), 
leading to
\bea
{\cL}_{\mbox{\tiny $m$II}} &=& \sqrtg \Bigl\{ e^{-2\phi} \Bigl[
            R\ -\ 4 (\part \phi)^2\ +\ \tfrac{1}{12} H_\mnr(B)H^\mnr(B) 
         \ +\ \tfrac{1}{8} \Tr (\part M^{-1} \part M) \nn  [.3cm]
&& \hsp{3cm}
          + \tfrac{1}{k^2}(\part k)^2 
         \ -\ \tfrac{1}{4}k^2 F_{\mn}(A)F^{\mn}(A)
         \ -\ \tfrac{1}{4k^2} F_{\mn}(B) F^{\mn}(B) \Bigr]    \nn[.3cm]
&&  \hsp{1cm}
         - \ \tfrac{1}{4} k F_{\mn}^a(V)\Mab  F^{b\mn}(V)
        \ + \ \tfrac{1}{2k} D_\mu\ell^a \Mab D^\mu\ell^b 
      \ - \ 2 k\ m^a \Mab m^b  \Bigl\} \nn  [.3cm]
&-&\tfrac{1}{8} \ \epsvijf \eta_{ab} \Bigr\{ 
    D_\mu \ell^a F^b_{\nu\rho}(V)( B_{\lambda\sigma} - A_\lambda B_\sigma )
    + 2 (V^a_\mu + \ell^a A_\mu) F^b_{\nu\rho}(V)  F_{\lambda\sigma}(B) 
\label{mII3}
\\[.3cm]
&& \hsp{6cm}   
   - 2  (V^a_\mu + \ell^a A_\mu) D_\nu \ell^a H_{\rho\lambda\sigma}
\Bigr\}\ .  \nonumber
\eea
After rewriting the topological terms, this action 
coincides with action (\ref{mII}), obtained from dimensionally reducing
the massive Type IIA theory. 

\sect{Massive T-duality rules in six dimensions}

Now that we reduced two theories \mIIA\ and IIB in six dimensions onto 
the same five-dimensional action $m$II, it is straight forward to 
construct a direct mapping between the unreduced theories. This mapping is 
the massive T-duality, a generalisation of the (massless) T-duality between
(massless) Type IIA and Type IIB \cite{Witten, DLR, BBJ}.  

Comparing the \mIIA\ and IIB reduction rules (\ref{mIIAred}) and 
(\ref{IIBred}), we find the following $m$T-duality rules (denoting the 
\mIIA\ fields with a prime): 
\be
\begin{array}{ll}
\hg_{xx}^\prime = 1/\hg_{xx},  
& \hB_{\mu x}^\prime  = \hg_{\mu x}/\hg_{xx}  ,
\\[.3cm]
\hg_{\mu x}^\prime  = \hA_{\mu x}/\hg_{xx},  
& \hB_{\mn}^\prime = \hA_{\mn} + 2 (\hg_{x [\mu}\hA_{\nu] x})/\hg_{xx} ,   
\\[.3cm]
\hg_{\mn}^\prime = \hg_{\mn} 
                - (\hg_{\mu x}\hg_{\nu x} - \hA_{\mu x}\hA_{\nu x})
                                     /{\hg_{xx}} ,  \hsp{1cm} 
& \hV^\prime{}_x^a = \hell^a + 2m^a x,
\\[.3cm]
\hphi^\prime = \hphi - \half \log |\hg_{xx}|, 
& \hV^\prime{}_\mu^a = \hV^a_{\mu x} + \hell^a \hA_{\mu x},
\\[.3cm]
{\hM^\prime}{}_{ab}^{-1} = \hMab,
\end{array}
\ee
and inversely (denoting the IIB fields with a prime):
\be
\begin{array}{ll}
\hg_{xx}^\prime = 1/\hg_{xx},  
& \hA_{\mu x}^\prime  = \hg_{\mu x}/\hg_{xx}  ,
\\[.3cm]
\hg_{\mu x}^\prime  = \hB_{\mu x}/\hg_{xx},  
& \hA_{\mn}^\prime = \hB_{\mn} + 2 (\hg_{x [\mu}\hB_{\nu] x})/\hg_{xx} ,   
\\[.3cm]
\hg_{\mn}^\prime = \hg_{\mn} 
                - (\hg_{\mu x}\hg_{\nu x} - \hB_{\mu x}\hB_{\nu x})
                                     /{\hg_{xx}} ,  \hsp{1cm} 
& \hell^{\prime a} = \hV_x^a - 2m^a x,
\\[.3cm]
\hphi^\prime = \hphi - \half \log |\hg_{xx}|, 
& \hV^\prime{}_{\mu x}^a = \hV^a_{\mu} -  ( \hV_x^a - 2m^a x)
                                          \hg_{\mu x}/\hg_{xx} ,
\\[.3cm]
{\hM^\prime}{}_{ab}^{-1} = \hMab,
\end{array}
\ee
The massive T-duality rules have extra massive corrections with respect to 
the massless ones in the transformation of the \Ovt gauge fields and 
scalars. Through the Scherk-Schwarz procedure these corrections have a 
linear dependence on the coordinate in which the T-duality is been 
performed.  
Note that we do not give explicitly the T-duality rules relating the 
$\hV_{\mn}^a$, $\hC_{\mu x}$ and  $\hC_{\mn}$ fields to Type IIA fields. 
It turns out they are non-locally related to $\hV^\prime{}_\mu^a$,  
$\hg^\prime_{\mu x}$ and $ \hB^\prime_{\mu x}$ 
respectively, as can be seen from (\ref{self-dual5}).

\sect{Heterotic theory and massive S-duality?}  

Another well-known duality between $N=2$ string theories is the 
string-string duality between the Heterotic string compactified on $T^4$ 
and Type IIA compactified on $K3$ \cite{Witten, HT, Sen, HS, DLR}. This 
is a 
non-perturbative duality, relating the strong coupling regime of one 
theory to the weak coupling regime of the other and vice versa.
  
There seems no straightforward generalisation of this duality to a 
massive string-string duality in six dimensions between massive IIA on 
$K3$ and Heterotic on $T^4$, one of the most obvious reasons being that 
the symmetry groups do not match. Heterotic on $T^4$ has an \Ovt\ symmetry
and performing a gauging analogous to (\ref{gauge}) would give a massive 
five-dimensional theory with $O(3,19)$ symmetry.

However, Heterotic theory on $T^5$ has an \Ove symmetry and performing a 
Scherk-Schwarz reduction by gauging the shift transformation (\ref{gauge}),
gives an \Ovt invariant massive $N=4$ gauged supergravity in four 
dimensions \cite{BRE}.
Therefore, naively one might think that a massive S-duality might be 
possible in five dimensions after a Scherk-Schwarz reduction of Heterotic 
on $T^5$ and the standard Kaluza-Klein reduction of $m$II theory, whose 
action was given in (\ref{mII}) and
(\ref{mII3}).

However a closer study reveals that such a massive S-duality does not work,
as pointed out in \cite{HLS}. There are various way to see this:
\begin{enumerate}
\item It turns out that, although the two theories have the same field 
content, the behaviour of the fields under the massive gauge 
transformations is different in each theory. Where the $m$II theory 
(\ref{mII}) the \Ovt vectors are St\"uckelberg fields that get eaten by 
the Kalb-Ramond form, in the Scherk-Schwarz reduction of the Heterotic 
theory it is the $V_\mu^a$ that become massive through the St\"uckelberg 
transformation of scalars coming from one of the \Ove vectors \cite{BRE}. 

\item Massless string-string-string triality \cite{DLR} between Heterotic 
theory on $T^4$, Type IIA on $K3$ and Type IIB on $K3$ tells us that the 
Heterotic theory on $T^5$ can be thought of as a (standard) Kaluza-Klein 
reduction of Type IIB. On the other hand, $m$II theory (\ref{mII}) can be 
thought of as a Scherk-Schwarz reduction of the same Type IIB. Since 
Scherk-Schwarz and Kaluza-Klein reduction do not commute, it is clear that 
the Scherk-Schwarz reduction of Heterotic theory  on $T^5$ will give a 
different result as the Kaluza-Klein reduction of the $m$II theory.      

\item The (massless) string-string-string triality in six dimensions works 
because in five dimensions there exists only one massless $N=2$ theory. 
Hence, all six-dimensional theories can be related to each other after 
mapping 
them onto the same five-dimensional theory. However this is no longer the 
case for massive theories. Since Scherk-Schwarz and Kaluza-Klein reduction 
do not commute, the number of inequivalent massive theories grows quickly 
as one reduces more and more, giving 15, 35 and 71 different theories in 
six, five and four dimensions respectively \cite{CLPST}. This in itself is 
of course not a proof for the impossibility of the massive S-duality, but 
at least it gives an idea of the large number of possibilities, in 
contrast to the massless case.         
\end{enumerate}

\sect{Conclusions}

We generalised the known T-duality rules between six-dimensional Type IIA 
and Type IIB compactified over $K3$ to massive T-duality between Romans' 
theory and Type IIB over $K3$. The obtained $m$T-duality rules recieve 
extra massive correction terms with linear dependence in the direction in 
which the $m$T-duality is done. This result is analogous to the 
ten-dimensional $m$T-duality rules given in \cite{BRGPT}.

It was shown in \cite{HLS} that the six-dimensional massive Type IIA theory
interpolates between ten-dimensional Romans' theory and massless Type IIA 
theory with non-trivial two- and four-form fluxes wrapped over the 
various two- and four-cycles. For example, using a particular symmetry of 
the \Ovt group, a D8-brane wrapped over the $K3$ can be dualised into a 
D4-brane solution of massless Type IIA on $K3$ with a four-form flux 
along the $K3$. 

Following the same line of thought, it is not difficult to see that this 
duality can be extended to include the Type IIB theory: the \Ovt group of 
the five-dimensional massive Type II theory we constructed above will map
solutions of Romans' theory into solutions of Type IIB with extra three- 
and five-form fluxes wrapped over the various non-contractable cycles of 
$K3 \x S^1$. In particular, the D8-brane wrapped over $K3 \x S^1$ will be 
dual to a D3-brane solution of Type IIB on $K3 \x S^1$ with an extra $F_5$
form field in the compactified space. 

So far, the generalisations of string dualities between massive 
supergravities have all been generalisations of the perturbative 
T-duality. As argued in \cite{HLS} and section 6, a massive version of 
the non-perturbative string-string duality is not possible in 
six-dimensions. The question arises whether massive S-dualities can be 
constructed for other cases or whether the mass terms turn out to be 
incompatible with non-perturbative dualities. If the latter turns out to be
the case, this adds another question to the mysterious features surrounding
(some of) the massive supergravity theories. In fact, the question of the 
strong coupling behaviour of (some of) these massive supergravity theories 
is related to the unknown eleven-dimensional origin of Romans' theory and 
its place in string theory. Still, as Romans' theory should be considered 
as the real effective low energy limit of Type IIA string theory, being the
only supergravity theory that gives rise to D8-brane solutions, one 
might expect that the U-duality group would, in some way or another, 
provide a non-perturbative duality for these supergravities.

\vspace{1cm}
\noindent
{\bf Note added}\\
Similar results on the six-dimensional massive T-duality rules will also 
soon appear in \cite{BBRS}.


\vspace{1cm}
\noindent
{\bf Acknowledgements}\\
The author is very grateful to Dominic Brecher, Clifford Johnson, 
Mees de Roo, Eduardo Eyras, Laur J\"arv, Patrick Meessen and Paul Saffin
for the useful discussions.


\end{document}